\input harvmac.tex
\input epsf
\let\includefigures=\iftrue
\includefigures
\input epsf
\epsfclipon
\def\fig#1#2{\topinsert\epsffile{#1}\noindent{#2}\endinsert}
\else
\def\fig#1#2{}
\fi
\def\Title#1#2{\rightline{#1}
\ifx\answ\bigans\nopagenumbers\pageno0\vskip1in%
\baselineskip 15pt plus 1pt minus 1pt
\else
\def\listrefs{\footatend\vskip
1in\immediate\closeout\rfile\writestoppt
\baselineskip=14pt\centerline{{\bf
References}}\bigskip{\frenchspacing%
\parindent=20pt\escapechar=` \input
refs.tmp\vfill\eject}\nonfrenchspacing}
\pageno1\vskip.8in\fi \centerline{\titlefont #2}\vskip .5in}
 
scaled\magstep3
 
scaled\magstep3
 
scaled\magstep3
 
scaled\magstep3
 
scaled\magstep3
\ifx\answ\bigans\def\tcbreak#1{}\else\def\tcbreak#1{\cr&{#1}}\fi
%
%

\def\inbar{\,\vrule height1.5ex width.4pt depth0pt}
\def\IB{\relax{\rm I\kern-.18em B}}
\def\IC{\relax\hbox{$\inbar\kern-.3em{\rm C}$}}
\def\IP{\relax{\rm I\kern-.18em P}}
\def\IR{\relax{\rm I\kern-.18em R}}

\def\sgn{{\rm sgn~}}

\def\b{\beta}

\def\D{\Delta}
\def\l{\lambda}

\def\barint{-\hskip -11pt\int}
\def\D{\Delta}
\def\sgn{{\rm sgn}}
\def\cut{\hskip 2pt{\cal /}\hskip -6.8pt}
\def\sn{{\rm sn}}
\def\cn{{\rm cn}}
\def\dn{{\rm dn}}
\Title{\vbox{\baselineskip12pt
\hfill{\vbox{
\hbox{LPTENS-95/56} \hbox{CERN-TH/95-352}
\hbox{hep-th/9601069\hfil}
}}}}
{\vbox{\centerline{Exact Solution of Discrete Two-Dimensional
$R^2$ Gravity }}}

\bigskip
\bigskip
\centerline{Vladimir A. Kazakov $^{(1)}$}
\smallskip
\centerline{Matthias Staudacher $^{(1),(2),\dagger}$}
\smallskip
\centerline{{\it and}}
\smallskip
\centerline{Thomas Wynter $^{(1),\dagger}$}\footnote~{
\hskip -11.5pt $^\dagger$ \hskip 4pt
This work is supported by funds provided by the European Community,
Human Capital and Mobility  Programme.}
\bigskip
\centerline{(1) Laboratoire de Physique Th\'eorique de}
\centerline{l'\'Ecole Normale Sup\'erieure\footnote*{
Unit\'e Propre du
Centre National de la Recherche Scientifique,
associ\'ee \`a l'\'Ecole Normale Sup\'erieure et \`a
l'Universit\'e de Paris-Sud.}}
\smallskip
\centerline{(2) CERN, Theory Division}

\bigskip
\noindent

We exactly solve a special matrix model of dually weighted planar
graphs
describing pure two-dimensional quantum gravity with an $R^2$
interaction. It permits us to study the intermediate regimes between the
gravitating and flat metric. Flat space is modeled by a
regular square lattice, while localised curvature is introduced
through lattice defects.
No ``flattening'' phase transition is found with respect
to the $R^2$ coupling: the infrared behaviour of the system is that
of
pure gravity for any finite $R^2$ coupling. In the limit of infinite
coupling, we are able to extract a scaling function interpolating
between pure gravity and a dilute gas of curvature
defects on a flat background.
We introduce and explain some novel techniques concerning our method
of
large $N$ character expansions and the calculation of Schur
characters on
big Young tableaux.

\Date{CERN-TH/95-352}

\nref\KSWI{V.A.~Kazakov, M.~Staudacher and T.~Wynter, \'Ecole Normale
preprint LPTENS-95/9, hep-th/9502132,
accepted for publication in
Commun.~Math.~Phys.}
\nref\KSWII{V.A.~Kazakov, M.~Staudacher and T.~Wynter, \'Ecole
Normale
preprint LPTENS-95/24, hep-th/9506174,
accepted for publication in
Commun.~Math.~Phys.}
\nref\POL{A.~Polyakov, Phys. Lett. 103B (1981) 207, 211.}
\nref\DAVID{F.~David, Nucl. Phys. B257 (1985) 45.}
\nref\VOL{V.A.~Kazakov, Phys. Lett. B150 (1985) 282.}
\nref\FRO{J.~Fr\" ohlich, in: Lecture Notes in Physics, Vol. 216,
Springer, Berlin, 1985; \hfill\break
J.~Ambj{\o }rn, B.~Durhuus and J.~Fr\" ohlich,
Nucl. Phys. B257[FS14](1985) 433.}
\nref\KKM{V.~Kazakov, I. Kostov, A. Migdal, Phys. Lett. 157B (1985)
295.}
\nref\K{V.~Kazakov, Phys. Lett. 119A (1986) 140.}
\nref\KPZ{V.~Knizhnik, A.~Polyakov and A.~Zamolodchikov, Mod. Phys.
Lett.
A3 (1988) 819.}
\nref\DDK{F. David, Mod. Phys. Lett. A3 (1988) 1651;
J. Distler and H. Kawai, Nucl. Phys. B321 (1989) 509.}
\nref\KAW{H.~Kawai and R.~Nakayama, Phys. Lett. 306B (1993) 224.}
\nref\ICH{S.~Ichinose, Nucl. Phys. B445 (1995) 311.}
\nref\KAZBOUL{D.V.~Boulatov and V.A.~Kazakov, Phys. Lett. 214B (1988)
581.}
\nref\BIPZ{E.~Brezin, C.~Itzykson, G.~Parisi and J.B.~Zuber, Commun.
Math. Phys. 59 (1978), 35.}
\nref\IDiF{P.~Di~Francesco and C. Itzykson, Ann. Inst. Henri.
Poincar\'e Vol. 59, no. 2 (1993) 117.}
\nref\BYRD{P.F.~Byrd and M.D.~Friedman, ``Handbook of Elliptic
Integrals for
Engineers and Physicists'', Springer, Berlin, 1954.}
\nref\LAWD{D.F.~Lawden, ``Elliptic Functions and Applications'',
Springer, New York, 1989.}
\nref\BUR{T.T.~Burwick, Nucl. Phys. B418 (1994) 257.}

\newsec{Introduction and overview}

Two-dimensional random geometry is now placed at the heart of many
models of modern physics, from string theory and two-dimensional
quantum
gravity, attempting
to describe fundamental interactions, to membranes and interface
fluctuations in various problems of condensed matter physics.
Therefore
it is quite important to study in detail the basic universal
properties of random geometries, such as their fractal nature, their
phase diagram, critical phenomena and correlations of physical
quantities of
geometric origin. In addition, one might hope to discover entirely
new
mechanisms for phase transitions.

Over the last fifteen  years, considerable progress has been achieved
in the
understanding of noncritical strings, or 2D quantum gravity in the
presence of matter fields with central charge $c \leq 1$. It was
based
on two different approaches, completing and justifying each other.

The first one, based on the continuous treatment of 2D metric
fluctuations by means of quantum Liouville theory,
was introduced by Polyakov \POL.

The second one, based on a discretisation of the two-dimensional
metric in terms of random planar graphs, was first proposed in
\DAVID,
\VOL\ and \FRO.

With the help of powerful matrix model techniques
this approach allowed for the first time the exact calculation of
critical exponents in quantum gravity;  both for pure 2D
gravity without matter \DAVID,\VOL, and, subsequently, for various
forms
of conformal matter with $c \leq 1$ (e.g.~$c=-2$ \KKM\ and
$c={1 \over 2}$ \K).
These results were successfully confirmed and supplemented by the
continuous approach in \KPZ, and then in \DDK.

Following this breakthrough,
numerous further developments have led to a rather
full understanding of the subject.

Still, we think that some important questions were
left  unclarified. In this paper, we are addressing one of them: what
will happen in the case of pure 2D gravity if we introduce
coupling constants favoring the flat configurations among the
ensemble of 2D metrics? Can we achieve a transition to a new phase of
essentially flat metrics? Or shall we always find -- on sufficiently
big
distances -- the familiar behaviour of 2D gravity, whatever these new
couplings are? If the last scenario is true, do we have some
universality for intermediate scales, when the size of the 2D
universe is of the order of the flattening scale?

On first sight, this question appears to have a simple answer, at
least from the point of view of the continuous theory.
The Euclidean path integral is
\eqn\cont{{\cal Z} = \int {\cal D}g_{ab}~e^{-\int d^2z \sqrt{\det g}
{}~(\mu + \alpha R_g + {1\over \b_0} R_g^2 + \ldots) } .}
Apart from the cosmological term (controlled by $\mu$)
and the (topological) Einstein term proportional to the curvature
$R_g$,
it does not seem to be meaningful to put further terms into the
action of 2D
gravity on dimensional grounds. The simplest term one might want to
consider
is ${1 \over \b_0} R_{g}^{2}$. The bare coupling constant $\b_0$ is
however
dimensionful and thus should be proportional to the cutoff squared.
So
it is small and in principle should be dropped, as well as any
further
higher derivative terms.

On the other hand, no argument has ever been given to exclude the
following alternative scenario: increasing
the $R_{g}^{2}$ coupling ($\beta_0 \rightarrow 0$ in our notation),
the
theory might entirely restructure {\it nonperturbatively}.
In this case the Liouville approach, which is known
to correctly quantise the theory \cont\ in the absence
(i.e.~$\b_0 = \infty$) of higher derivative terms, might
no longer yield a good description of the theory. Some calculational
attempts to address this question
within the framework of Liouville theory are inconclusive precisely
because they merely perturb by the $R^2$ term in question
(see e.g.~\KAW,\ICH).

In this paper, we address and resolve some of these questions in a
completely nonperturbative way. We take as a starting point
the by now very familiar discrete definition \DAVID,\VOL,\FRO\
of the path integral \cont,
known to correctly quantise the theory in the absence
of higher derivative terms. We then introduce couplings whose naive
continuum limit precisely induces the higher terms indicated
in \cont. Technically, this is done by solving an
unusual matrix model that generates the
dually weighted graphs proposed and studied in our earlier papers
\KSWI\ and \KSWII.

The paper is organized in the following way:

In section 2, the discretised model of dually weighted graphs is
formulated and the main physical results, including the absence of
a ``flattening'' phase transition, the joint scaling function of the
$R^2$
coupling and the cosmological constant and their infrared and
ultraviolet asymptotics are presented and discussed in a
non-technical
fashion.
We physically interpret our model as a
statistical mechanics system of a ``gas'' of point-like curvature
defects
surrounded by flat two-dimensional space.
Curvature is quantised, i.e.~localised on defects whose deficit angle
equals an integer times some fixed value.

In section 3, a review of the character expansion approach to the
models of dually weighted graphs (worked out in \KSWI\ and \KSWII) is
given.

In section 4, the exact solution of the present model is derived in
some
detail. The solution is rather complex but nevertheless explicit.
Elliptic functions, already crucial in \KSWII, play a central role.

Section 5 demonstrates how to extract the physical results from
the involved expressions of section 4.

Section 6 is devoted to conclusions and open questions.
We also discuss the important open issue of the
universality of our results and point out that fine-tuning the
curvature weights -- an in principle feasible extension of the
present work
-- might result in new phases of 2D gravity.

Finally, appendix 1.~contains the calculation of the Schur characters
of $GL(N)$ in the limit of a big Young tableau while some more
technical
details concerning the exact solution and its verification
are given in appendix 2.

\newsec{Definition of the model and physical results}

In the particular model used here
the two-dimensional random geometry is described
by planar graphs (with the topology of the sphere) built from
plaquettes
(``quadrangulations''). We are summing over all such graphs with the
weights
$t_2,t_4,t_6,....$, corresponding to vertices with
2,4,6,... neighboring plaquettes, respectively. The
partition function is
\eqn\DWGG{
{\cal Z}=
\sum_{G} \prod_{v_{2q} \in G}  {t_{2q}}^{ \# v_{2q}},}
where $v_{2q}$ are the vertices with coordination number $2q$
and $\# v_{2q}$ are the numbers of such
vertices in the given graph $G$. Note that one cannot sum up these
graphs by the usual one-matrix model, since we need to control
both vertex and face coordination numbers.
The discrete, curved manifold is thus described by a graph made from
flat squares. The curvature appears exclusively in the
vertices and corresponds to the deficit angles around the vertices:
$\pi (2-q)$. Thus two neighboring plaquettes contribute positive
curvature, four correspond to zero curvature and six and more result
in
localised negative curvature (see Fig~2.1).

\vskip 20pt
\epsfbox{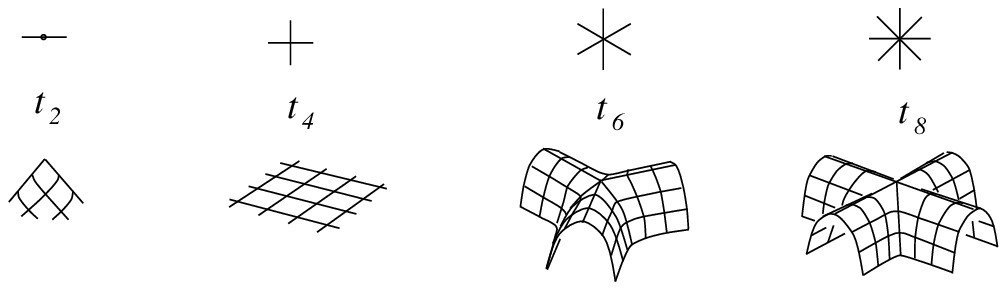}
\centerline{{\bf Fig.~2.1} Flat space and curvature defects.}
\vskip 30pt

For technical reasons, it proved useful to choose a
particular parametrisation of the above weights $t_{2 q}$:
\eqn\weights{
t_2=\sqrt{\l}~t,~t_4=\l,~t_6=\l^{3 \over2}~{\b^2 \over t},~...~
t_{2q}=\l^{q \over 2}~({\b^2 \over t})^{(q-2)}.}
With these weights, it is easy to prove, using Euler's theorem, that
the partition sum \DWGG\ becomes
\eqn\PART{
{\cal Z}(t,\l,\b) = t^4~\sum_{G}~\l^A~\b^{2 (\# v_{2}-4)}, }
where $A$ is the number of plaquettes of the graph $G$ and $\#v_2$
the
number of positive curvature defects. Note that the latter are
balanced by a gas of negative curvature defects, whose individual
probablities are given in \weights.

We expect this model to describe pure gravity in a
sufficiently large interval of $\b$, after tuning
the bare cosmological constant $\l$ (controlling the number
of plaquettes) to some critical
value $\lambda_c(\b)$. On the other hand, for $\l$ fixed and $\b=0$
we
entirely suppress curvature defects except for the four positive
defects
needed to close the regular lattice into a sphere. It is thus clear
that $\b$ is the precise lattice analog of the bare curvature
coupling
$\b_0$ in \cont.
The phase $\b=0$
of ``almost flat'' lattices -- very different from pure gravity --
was discussed in detail in \KSWII.

Let us now summarize the main physical results following from the
exact solution (for general $\l$ and $\b$) of this model:

1. A long debated question was whether models of the present type
undergo a ``flattening'' phase transition at a finite, non-zero
critical value of $\b=\beta_c$. The weak coupling region $\b >
\beta_c$
would then correspond to the standard phase of pure gravity while
a putative novel ``smooth'' phase of gravity might exist either
at $\b = \beta_c$ or in the entire interval $0 \leq \b \leq \beta_c$.
This would constitute an existence proof of {\it continuum}
2D $R^2$ gravity.
We find, to the contrary, that
{\it there is no ``flattening'' phase transition at non-zero
$\b$}. For any given $\b$ we find the powerlike
scaling of standard pure gravity on large scales.
This means that no matter how flat the system is on small scales (of
the order
of $\beta^{-{1 \over 2}}$), it destabilizes in the infrared
into the familiar ensemble of highly fractal ``baby-universes''.

2. The dependence of the partition sum \PART\
on $\b$ and the
lattice cosmological constant $\lambda$ in the vicinity
of the flat phase $\beta \sim 0$ and close to
$\l \sim \lambda_c$ is given by a simple,
(presumably) universal scaling function $f(x)$
(defined through ${\cal Z}(t,\l,\b) = {4 t^4\over 15\beta^2}~f(x)$)
reflecting the transition
from flat space to pure gravity:
\eqn\scalf{
f(x)=x^6 - {5 \over 2}~x^4 + {15 \over 8}~x^2 - {5 \over 16} -
     x~\big(x^2-1\big)^{5 \over 2},}
where the scaling variable $x$ is given, to leading order, by
\eqn\scalx{ x={\sqrt{2} \over \pi}~{1-\lambda \over \beta}.}
We can distinguish the following features:

(a) There is a degree ${5 \over 2}$ singularity at $x=1$, correctly
reproducing the universal string susceptibility exponent
$\gamma_s = -{1 \over 2}$ of pure gravity \DAVID,\VOL.
In view of eq.\scalx, the critical value of the lattice cosmological
constant $\lambda$ is therefore given to leading order by
$\lambda_c=1-{\pi \over \sqrt{2}} \beta + {\cal O}(\b^2)$.
Therefore, in view of \PART,
the characteristic growth of the random surfaces as
a function of the lattice area $A$ (= number of plaquettes)
is given by
\eqn\growth{{\cal Z}(t,A,\b)~\sim~{t^4\over\beta^{9\over 2}}~e^{{\pi
                           \over \sqrt{2}}~\beta~A}~A^{-{7 \over 2}}.
}
For any non-zero $\beta$ we do have exponential growth of the
number of surfaces, but one has to go to larger and larger scales
(i.e.~use more and more plaquettes) to be able to take the continuum
limit.
If $\beta$ is exactly zero there is no longer any exponential growth
and
no pure gravity continuum limit is possible. The prefactor
$\beta^{- {9 \over 2}}$ is found in the exact calculation in
section 5; we are not sure whether it is universal.

(b) We further see that taking $\beta \rightarrow 0$ {\it before}
the limit $\lambda \rightarrow \lambda_c$ corresponds to the limit
$x \rightarrow \infty$. In this limit one finds
$f(x) \sim {5 \over 128}~{1 \over x^2} + {\cal O}({1 \over x^3})$,
that is, the characteristic critical behavior of 2D gravity
``silently'' disappears and we recover a power series in
${1 \over x}$ corresponding, in view of \scalx,
to a perturbative expansion in lattice defects $\beta$. In this limit
the
characteristic growth of surfaces as a function of area $A$ is
\eqn\limit{{\cal Z}(t,A,\b)~\sim~t^4~(~A+{\cal O}(\beta^2 A^3)~).}
The leading order corresponds precisely to the almost flat
lattices (with exactly four positive defects) of \KSWII.
The corrections are interpreted as insertions of negative
defects, balanced by further positive defects.
The typical shape of the surfaces in this limit is a generalisation
of the one we found for ``almost flat'' graphs in \KSWII:
Long, filamentary cylinders growing out from every negative
curvature defect.

(c) It is easy to prove that the scaling function \scalf\ is
the simplest possible function with the limiting properties
discussed in (a) and (b).

The above results might be interpreted in terms of
a continuum model of quantised curvature defects, in which
the localised defects move around like particles in a gas
on a flat background space.
The deficit angle, $\D \theta$, of a defect can take on the
values $\D \theta =\pi, 0$ and $-\pi$. A positive curvature defect is
surrounded by a conical geometry, whereas a negative curvature defect
corresponds to a saddle-type insertion (see Fig.~2.1).
The higher order negative
curvature defects ($-2\pi$ and higher) would not be expected to play
a
role in this limit (the entropy from moving two low order defects
around would completely dominate that from a single higher order
defect).  The coupling $\beta$ can be interpreted as a fugacity
controlling the number of defects. The flat space limit
$\beta\rightarrow 0$ consists of four defects of degree $\pi$ moving
around
with respect to one another. Varying the fugacity, $\beta$, allows
one
to smoothly interpolate between flat space, \limit\ (with four
defects), and pure gravity \growth\ (with an infinite number of
defects).

\newsec{Review of the technique}

The appropriate techniques for dealing with models of dually weighted
planar graphs were pioneered in two previous publications
\KSWI,\KSWII.
Let us recall the key steps of the method and collect the crucial
formulas allowing the exact solution of the model.
The method is in fact applicable to a situation more general than the
one
written in eq.\DWGG, to be analysed
in the present work. The partition function of general
dually weighted planar graphs $G$ is defined to be
\eqn\DWG{
Z(t^*,t)=
\sum_{G} \prod_{v^*_q,v_q \in G} {t_q^*}^{\# v^*_q}\ {t_q}^{ \#
v_q},}
where $v_q^*,v_q$ are the vertices with $q$ neighbours on the
original
and dual graph, respectively, and $\# v_q^*,\# v_q$ are the numbers
of
such vertices in the given graph $G$. This expansion is
generated by the following matrix model:
\eqn\DWGmatrix{
Z(t^*,t)=\int\,{\cal D}M\ e^{-{N\over 2} \Tr~M^2\ +\ \Tr~V_B(M A)},}
with
\eqn\potential{
V_B(M A)=
\sum_{k=1}^{\infty}{1\over k}~\Tr B^k\ (M A)^k .}
The matrices $A$ and $B$ are fixed, external matrices encoding the
coupling constants through
\eqn\tqAB{
t_q^*={1\over q}\ {1\over N}\ \Tr\ B^q
{\rm \hskip 20pt and \hskip 20pt}
t_q={1\over q}\ {1\over N}\ \Tr\ A^q.}
The model generalises, for $A \neq 1$, the standard one matrix model
first solved by Br\'ezin, Itzykson, Parisi and Zuber \BIPZ.
It may no longer be successfully treated by changing to
eigenvalue variables; a reduction to $N$ variables is nevertheless
possible:
One expands the potential in terms of
the characters on the group:
\eqn\potchar{
e^{\Sigma_{k=1}^{\infty} { 1 \over k} \Tr B^k\ \Tr (MA)^k}=
\prod_{i,j=1}^N{1\over (1-B_i(MA)_j)}
={1\over N^N}\sum_R\chi_R(B)\ \chi_R(MA).}
Here $B_i$ and $(MA)_j$ are the eigenvalues of the matrices $B$ and
$MA$. The first step involves rewriting the sum over $k$ as a double
sum over all the eigenvalues of the matrices $B$ and $MA$ of
$-\ln(1-B_i(MA)_j)$. Exponentiating the log then gives the product
in the numerator. The second step uses
a group theoretic result to write the product in terms
of a sum over characters. The character is defined by the Weyl
formula:
\eqn\eigchar{
\chi_{\{h\}}(A)={\det_{_{\hskip-2pt (k,l)}}(a_k^{h_l})\over
\Delta(a)},}
where the set of $\{h\}$ are a set of ordered, increasing,
non-negative integers, $\D(a)$ is the Vandermonde determinant, and
the
sum over $R$ is the sum over all such sets.  The $R$'s label
representations of the group $U(N)$ and the sets of integers $\{h\}$
are the usual Young tableau weights defined by
$h_i=i-1+\#$boxes in row $i$ (the index $i$ labels the rows in
the Young tableau, $i=1$ corresponding to the first row). Note
that the restriction on the allowed Young tableaux that any row must
have at least as many boxes as the row below implies that the
$\{h_i\}$ are a set of increasing integers:
\eqn\hconstr{
h_{i+1}>h_i.}
Substituting
equation \potchar\  into the integral in equation \DWGmatrix, we can
now do the angular integration using the key identity
\eqn\angular{
\int\,({\cal
D}\Omega)_H~\chi_R(\Omega M\Omega^{\dagger}A)=
d_R^{-1}~\chi_R(M)~\chi_R(A),}
where $d_R$ is the dimension of the representation given by
$d_R=\D(h)/\prod_{i=1}^{N-1}i!$), and arrive, after performing
a Gaussian integral over the eigenvalue degrees of freedom,
at the expression
\eqn\IzDiFr{
Z(t,t^*)=c\,\sum_{\{h^e,h^o\}}
{\prod_i(h^e_i-1)!!h^o_i!!\over
\prod_{i,j}(h^e_i-h^o_j)}~\chi_{\{h\}}(A)~\chi_{\{h\}}(B).}
The sum is taken over a subclass of so-called even representations.
These are defined as possessing an equal number of even weights
$h^{e}_{i}$ and odd weights $h^{o}_{i}$ (since the mentioned Gaussian
integration vanishes if the latter condition is not satisfied).
The formula \IDiF\ was originally discovered by Itzykson and Di
Francesco
\IDiF\ by summing up ``fatgraphs'', using purely combinatoric
and group theoretic arguments.

In our first paper \KSWI\ we demonstrated
how to take the large $N$ limit of this
expansion. In this limit the weights ${1 \over N} h_i$ condense to
give a smooth, stationary distribution $dh~\rho(h)$, where $\rho(h)$
is a probability density normalized to one.  For simplicity we
restrict our attention to models in which the matrices $A$ and $B$
are such that traces of all odd powers of $A$ and $B$ are zero. This
means that the our random surfaces are made from vertices and faces
with even coordination number. As was discussed in \KSWI, this
ensures
that the support of the density $\rho(h)$ lies entirely on the real
axis and thus simplifies the solution of the problem.
The matrix $A$ will satisfy this condition if we
introduce a ${N\over 2}\times{N\over 2}$ matrix $\sqrt{a}$ in terms
of which $A$ and the character $\chi_{\{h\}}(A)$ are given by
\eqn\Aa{
A=\left[\matrix{\sqrt{a}&0\cr 0&-\sqrt{a}\cr}\right]\quad {\rm and}
\quad
\chi_{\{h\}}(A)=
    \chi_{\{{h^e\over 2}\}}(a)\chi_{\{{h^o-1\over 2}\}}(a)
    \,\,\sgn\bigl[\prod_{i,j}(h^e_i-h^o_j)\bigr].}
In order to study the transition from flat to random graphs we choose
the potential to be

\eqn\models{
V_A(MA_4)
=\sum_{k=1}^\infty{1\over 2k}~\Tr[A^{2k}]~(MA_4)^{2k}. }
Here $A_4$ is defined to satisfy $\Tr[(A_4)^k]=N \delta_{k,4}$
and $A$ is as defined in \Aa. We are thus studying surfaces
made up entirely from squares. A weight $t_{2 k}=Tr[A^{2 k}]$ is
assigned
whenever $2 k$ squares meet at a vertex (see Fig.~2.1).
We are therefore precisely considering the situation defined in
eq.\DWGG.
In order to investigate \IzDiFr\ in the large $N$ limit,
one attempts to locate the stationary point.
This leads to the saddlepoint equation
\eqn\sdpt{
2F(h)+\barint_0^a\ dh'\ {\rho(h') \over h-h'}= -\ln h.}
As has been discussed in detail in \KSWI, this equation actually does
not hold on the entire interval $[0,a]$, but only on an interval
$[b,a]$ with $0 \leq b \leq 1 \leq a$:
Assuming the equation to hold on $[0,a]$ would violate the implicit
constraint $\rho(h) \leq 1$ following from the restriction \hconstr.
The density is in fact exactly saturated at its maximum value
$\rho(h)=1$
on the interval $[0,a]$.
The solution of \sdpt\ requires the knowledge of the large $N$ limit
of
the variation of the characters in eq.\Aa:
\eqn\defF{
F(h_k)=2{\partial \over \partial h^e_k}~\ln\ {\chi_{\{{h^e\over
2}\}}(a)
\over \Delta(h^e)}.}
A rather general method for its determination has been one of the
main
technical achievements of our previous work. In fact, we found the
following
simple result: Introduce the character function $F(h)$ and the
resolvent
$H(h)$ as, respectively, the large $N$ limit of \defF\ and
\eqn\res{H(h)=\int_0^a dh'\ {\rho(h') \over h-h'}.}
We found the weight resolvent $H(h)$ to be very closely related to
the
standard matrix model eigenvalue resolvent. It provides
a direct link between the statistical distribution
of Young weights and the correlators of the model:
\eqn\trMtwoq{
{1\over N}~\Tr M^{2q}=
{\lambda^q\over q}\oint\,{dh\over 2\pi i}~h^q~e^{qH(h)}.}
Here the contour encircles the cut of $H(h)$.
Further introduce the function $G(h)$ as
\eqn\defG{
G(h) = e^{H(h) +F(h)}.}
Its importance stems from the fact that it relates in
a simple way the introduced functions
$F(h)$,$H(h)$ and the coupling constants $t_{2 q}$ 
\KSWI:
\eqn\tqQ{
t_{2q}={1\over q}\oint\,{dh\over 2\pi i}~G(h)^q.}
Using this equation as well as an alternative representation
of the Weyl character \eigchar
\eqn\schuchar{
\chi_{\{h\}}(A)=\det_{_{\hskip -2pt (k,l)}}
                 \bigl(P_{h_k+1-l}(\theta)\bigr)}
as a determinant of Schur polynomials $P_n(\theta)$ defined through
\eqn\schupol{
e^{\Sigma_{i=1}^{\infty}z^i\theta_i}=\sum_{n=0}^{\infty}z^n
P_n (\theta)\quad {\rm with}\quad \theta_i={1\over i}~\Tr[A^i],}
one derives \KSWII\ from the ``equation of motion''
\eqn\difpoly{
{\partial\over\partial\theta_q}P_n(\theta)=P_{n-q}(\theta)\quad
{\rm with}\quad \theta_q={N\over 2q}t_{2q}}
the equation
\eqn\hGexp{
h-1 = \sum_{q=1}^Q{t_{2 q}\over G^q} +
\sum_{q=1}^{\infty}
{2q\over N}{\partial\over \partial t_{2q}}\ln\Bigl(
\chi_{\{{h^e\over 2}\}}(a)\Bigr)
{}~G^q,}
where the coefficients of the positive powers of $G$ in \hGexp\
are directly related to the correlators of the matrix model
dual to \models, i.e.~the model with potential
$V_{A_4}(\tilde M A) ={1\over 4}~(\tilde M A)^4$:
\eqn\dualcorr{
{2q\over N}{\partial\over \partial t_{2q}}\ln\Bigl(
\chi_{\{{h^e\over 2}\}}(a)\Bigr)=
\langle {1\over N}\Tr~(\tilde M A)^{2q} \rangle.}
Here the negative powers of $G$ are fixed by
eq.\tqQ\ while the positive powers are determined through
eq.\difpoly.
If we write formula \hGexp\ in the form
$\hat{h}~\chi_{h}(t)=
(\sum_{q \geq 0}~G^{-q}~t_{2 q} + \sum_{q > 0}
{}~G^q~{2 q \over N} {\partial \over \partial t_{2 q}})
{}~\chi_{h}(t)$
we see that the operator $\hat{h}$ acts on the character like an
operator of
the derivative of the bosonic field ${d \over dG}~\phi(G)$ since the
commutation relations of $\phi(G)$ for different $G$'s
are those of a bosonic field\foot{We acknowledge several
conversations
with I.~Kostov concerning this point.}.
This fact suggests that the models of dually weighted planar graphs
might have a description in terms of
integrable hierarchies of differential
equations, just as for the (much simpler) case of ordinary matrix
models.

We have also assumed for the moment that only a finite number $Q$
of couplings are non-zero (i.e.~$t_{2 q}=0$ for $q>Q$).
Furthermore, we were able to show in our second work \KSWII\
that \hGexp\ implies the functional equation
\eqn\HGprod{
e^{H(h)}={(-1)^{(Q-1)}h\over t_Q}\prod_{q=1}^Q G_q(h),}
where the $G_q(h)$ are the first $Q$ branches of the multivalued
function
$G(h)$ defined through \hGexp\ which map the point $h=\infty$ to
$G=0$. The saddlepoint equation \sdpt, together with \HGprod,
defines a well-posed Riemann-Hilbert problem. It was solved exactly
and
in explicit detail in \KSWII\ for the case $Q=2$, where the
Riemann-Hilbert
problem is succinctly written in the form
\eqn\twofph{\eqalign{
2\cut F(h)+H(h)=&-\ln (-{h\over t_4})\cr
2F(h)+\cut H(h)=&-\ln h,}}
where $\cut H(h)$ denotes the real part of
$H(h)$ on the cut $[b,a]$ and $\cut F(h)$ denotes the
real part of $F(h)$ on a cut $[-\infty,c]$ with
$c<b$. This case corresponds,
in view of the potential \models, to an ensemble of squares being
able to meet in groups of four (i.e.~flat points) or two
(i.e.~positive curvature points). We termed the resulting surfaces
``almost flat''. It turned out that all the introduced functions
could be found explicitly in terms of elliptic functions. We also
wrote down
the Riemann-Hilbert problem for the more complicated case
$Q=3$. Allowing in addition a non-zero negative curvature coupling
$t_6$, it already captures the transition from flat to random
surfaces. The explicit equations are
\eqn\grav{\eqalign{
2F(h) + \cut H(h) =& -\ln h\cr
F_1(h)+F_2(h)+F_3(h)+2H(h)=&-\ln({h\over t_6}).}}
Its exact solution can, in principle, be obtained as well.
Some of the steps of the solution of \grav\ as well as for the
general
case of an arbitrary number of couplings will be presented in
appendix 1.
Unfortunately it involves already functions more general than
elliptic,
making the solution much less explicit and transparent.
Luckily, however, we are not forced to solve the system \grav\ in
order
to analyse the problem of the transition from flat to random
lattices.
Indeed, we merely need to perturb our almost flat lattices by {\it
any}
operator containing negative curvature. This physical observation
will be used in the next section to achieve the long-sought solution
to our problem.

\newsec{Exact solution of discrete two-dimensional $R^2$ gravity}

In the review of the method of the previous section we recalled that
the analytic structure of the multivalued function $G(h)$ \defG\ is
in
part determined by the weights $t_{2q}$. Specifically, the structure
of
the physical sheet and of all the sheets attached to it
by the cut of $e^{F(h)}$ is the same as that given by (i) dropping
the
positive powers of $G$ in \hGexp\ and (ii) inverting this truncated
equation
to obtain $G$ as a function of $h$. In \KSWII\ this observation
permitted us to solve the flat space limit. There the equation
\hGexp\
reads:
\eqn\hGflat{
h-1={t_2\over G}+{t_4\over G^2}+{\rm\ positive\ powers\ of\ } G.}
Dropping the positive powers of $G$ leaves us with a quadratic
equation for $G$. We thus deduced that
the physical sheet and all the sheets attached to it
by the cut of $e^{F(h)}$ have a quadratic structure. In
other words, there is only one sheet attached by the cut of
$e^{F(h)}$. This simple structure allowed us to solve the flat space
limit directly.

The key observation that allows us to solve the full problem is that
we can choose a more general set of weights, including those needed
to
introduce negative curvature defects, whilst still preserving this
simple two sheet structure. Labeling the first two weights as in
\KSWII\ :
$t_2$, $t_4$, we now include all the even $t_{2q}$ with $q\geq 3$,
assigning them the following weights: $t_{2q}=t_4\epsilon^{q-2}$.
Equation \hGexp\ can then be written compactly as:
\eqn\hGfull{
h-1={t_2\over G}+{t_4\over G~(G-\epsilon)}+{\rm\ positive\ powers\
of\ } G.
}

Dropping the positive powers of $G$ and inverting this equation, we
see that there are two sheets connected together by a square root cut
running between two finite cut points, $d$ and $c$ (see Fig.~4.1). On
the physical sheet a further cut (running from $b$ to $a$),
corresponding to $e^{H(h)}$, connects to further sheets.

\vskip 20pt
\hskip 30pt \epsfbox{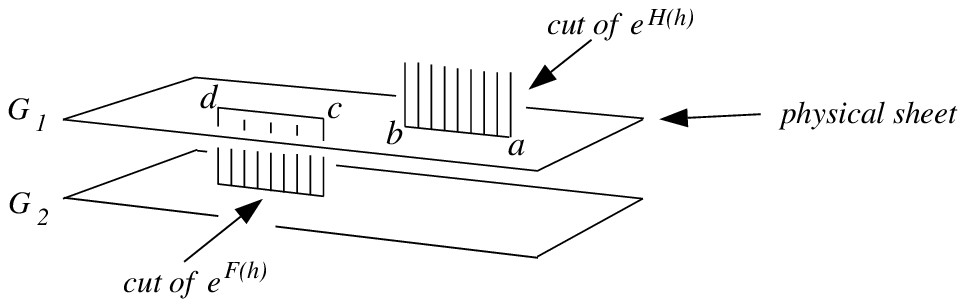}

\centerline{{\bf Fig.~4.1} Sheet and cut structure of $G(h)$ }
\vskip 30pt

\hskip -19pt To proceed
with the solution we relate the function $G(h)$ to the resolvent
$H(h)$ by the formula:
\eqn\HGprodmod{ e^{H(h)}={h\over \epsilon t_2-t_4}~G_1(h)~G_2(h),
}
where $G_1(h)$ and $G_2(h)$ are the physical sheet and the only other
sheet attached to it by the cut of $e^{F(h)}$ (see Fig.~4.1). Note
that
this differs
slightly from the form \HGprod\ derived in \KSWII\ since there are
now
an infinite number of inverse powers of $G$ in \hGexp . It can be
derived by the methods used in \KSWII .

We now proceed exactly as in \KSWII\ .
Equation \HGprodmod\ can be written
as
\eqn\fpfph{
F_1(h)+F_2(h)+H(h)=\ln ({h\over \epsilon t_2-t_4}),
}
where $\ln G_i(h)=F_i(h)+H(h)$. The two sheets $G_1(h)$ and
$G_2(h)$ are glued together by the square root cut coming from
$F(h)$. The combination $F_1(h) + F_2(h)$, evaluated on the cut of
$F(h)$, is twice the constant part of $F(h)$ on the cut (the
discontinuous part of $F(h)$ is of opposite sign on $F_1(h)$ and
$F_2(h)$ and is therefore canceled). We thus have the two equations
\eqn\twofphmod{\eqalign{
2\cut F(h)+H(h)=&-\ln ({h \over \epsilon t_2-t_4})\cr
2F(h)+\cut H(h)=&-\ln h,}}
the first coming from the large $N$ limit of the character \HGprodmod
and the second, in view of \res,
being the saddlepoint equation \sdpt. These two
equations tell us about the behaviour of the function $2F(h)+H(h)$ on
the cuts of $F(h)$ and $H(h)$, respectively. We have introduced the
notation $\cut F(h)$ to denote the real part on the cut of $F(h)$,
and
similarly for $\cut H(h)$. The principal part integral in \sdpt\ is
thus denoted in \twofphmod\ by $\cut H(h)$.

Our object is to now reconstruct the analytic function
$2F(h)+H(h)=2\ln G(h)-H(h)$ from its behaviour on its cuts. To do
this
we first need to understand the complete structure of cuts. We
already know
the structure of cuts of $H(h)$; it has a logarithmic cut running
from
$h=0$ to $h=b$, corresponding to the portion of the density which is
saturated at its maximal value of one, and a cut from $b$ to $a$
corresponding to the ``excited'' part of the density, where the
density is less than one. It thus remains for us to understand the
cut
structure of $\ln G(h)$.

The function $G(h)$ has two cuts on the physical sheet. The first
cut,
running from $b$ to $a$, corresponds to the cut of $e^{H(h)}$, the
second cut, running from cut point $c$ to cut point $d$, corresponds
to the cut of $e^{F(h)}$ (see Fig.~4.1). To see whether $\ln G(h)$
has any
logarithmic cuts, we first notice from \hGfull\ that $G(h)$ is non
zero
everywhere in the complex $h$ plane except possibly at infinity. Thus
for $\ln G(h)$ the only finite logarithmic cut points are at $h=b$,
defined to be the end of the flat part of the density (this
corresponds to the end of the cut of $e^{H(h)}$), and possibly the
cut
point $c$, defined to be at the end of the cut of $e^{F(h)}$. The
only
remaining question is whether this logarithmic cut starting at $h=b$
goes off to infinity or terminates at $c$. For large $h$ we see from
\hGfull\ that $\ln G(h)=\ln\epsilon+{\cal O}({1\over h})$, i.e. there
is no logarithmic cut at infinity. We conclude that the cut structure
of
the function $\ln G(h)$ consists of the cuts corresponding to
$e^F(h)$ and $e^{H(h)}$ connected together by a logarithmic cut,
whose cut points are $b$ and $c$ (see Fig.~4.1).

We now introduce two functions $\tilde F(h)$ and $\tilde H(h)$
defined by
\eqn\tildefh{
F(h) = \tilde F(h)-\ln {h\over (h-c)}\quad{\rm and}\quad
                 H(h) = \tilde H(h) + \ln {h\over h-b}.}
They are defined such that they have no logarithmic cuts.
The equation \twofphmod\ now reads
\eqn\twoftpht{\eqalign{
2\cut \tilde F(h)+\tilde H(h)=&\ln (t_4-\epsilon t_2)+
\ln\bigl(-{(h-b)\over (h-c)^2}\bigr)\cr
2\tilde F(h)+\cut \tilde H(h)=&\ln \bigl({(h-b)\over
(h-c)^2}\bigr).}}
These two equations define the behaviour of $2\tilde F(h) + \tilde
H(h)$ on all of its cuts. By standard methods we now generate the
full
analytic function $2\tilde F(h) + \tilde H(h)$. There are four cut
points, $a$ and $b$ defining the cut of $\tilde H(h)$, and $c$ and
$d$ defining the cut of $\tilde F(h)$. Their values will be fixed
later by boundary conditions. Note that this differs from the flat
space limit solved in \KSWII\ in that $\tilde F(h)$ now has two
finite
cut points whereas previously it had a semi-infinite cut
(i.e. $d\rightarrow -\infty$).  We generate
the full analytic function by performing the contour integral
\eqn\contint{
2\tilde F(h)+\tilde H(h)=r(h)\biggl[
\oint_{C_H}\,{ds\over 2\pi i}
          {\ln \bigl({(s-b)\over (s-c)^2}\bigr) \over (h-s)~r(s)}+
\oint_{C_F}\,{ds\over 2\pi i}
     {\ln (t_4-\epsilon t_2)+\ln\bigl(-{(s-b)\over (s-c)^2}\bigr)
\over
(h-s)~r(s)}\biggr],
}
where for compactness we have defined
\eqn\defr{
r(h)=\sqrt{(h-a)(h-b)(h-c)(h-d)}.
}
The contour $C_H$ encircles the cut $[b,a]$, whilst $C_F$
encircles the cut between $c$ and $d$.
They are illustrated in Fig.~4.2(a).

\vskip 20pt
\epsfxsize=450pt\epsfbox{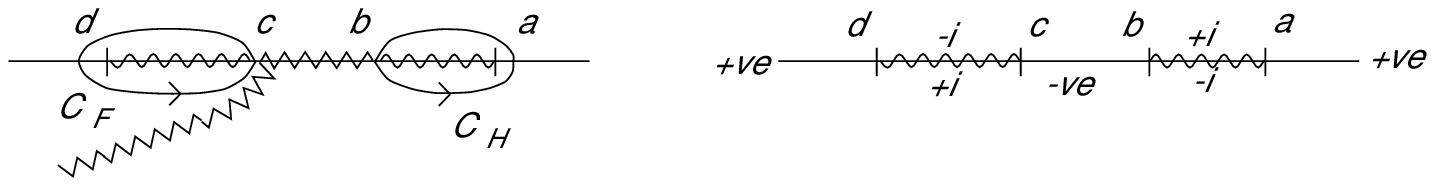}

\centerline{{\bf Fig.~4.2}\ \ (a)Contours $C_H$ and $C_F$.\quad\quad
(b)Sign convention for square roots of $r(h)$.}
\vskip 30pt

The zigzag line between $c$ and $b$ corresponds to the cut of
$\ln\bigl({(h-b)\over (h-c)}\bigr)$, The remaining zig-zag line
starting at $c$ corresponds to the remaining logarithmic cut,
$-\ln(h-c)$.
Expanding the contours, catching poles on the way and using the fact
that
logarithmic cuts have a discontinuity of $\pm i\pi$, we arrive at
\eqn\cbbainfd{\eqalign{
2F(h)+H(h)=&\ln {t_4-\epsilon t_2\over h}\cr &+ r(h)\biggl[
\int_c^b\,{ds\over (h-s)~r(s)}+
\int_b^a\,{ds\,{1\over \pi i}\ln(t_4-\epsilon t_2)\over
(h-s)~r(s)}-
\int_{-\infty}^d\,{ds\over (h-s)~r(s)}
\biggr].}}
Fig.~5(b) clarifies the sign convention for the square root
$r(h)$ on the real axis above and below the cuts.
Note that, for the cuts of $1/r(h)$  the signs on the
cuts are inverted compared to Fig.~5(b),
i.e.~$+i\leftrightarrow -i$. The integrals in \cbbainfd\ are
defined to be along the upper side.

To fix the constants $a$, $b$, $c$ and $d$, we expand \cbbainfd\ for
large $h$ and compare the resulting power series expansion to that
obtained from inverting \hGexp :
\eqn\bcexpn{
2F(h)+H(h)=2\ln\epsilon+\bigl({2t_4\over\epsilon^2}-1\bigr){1\over h}
\quad+\quad{\cal O}\bigl({1\over h^2}).
}
The terms of ${\cal O}\bigl({1\over h^2})$ depend on the as yet
unknown positive powers of $G$ in \hGfull . Expanding \cbbainfd\ for
large $h$ and comparing to \bcexpn\ we find the three boundary
conditions:
\eqn\bcone{
\ln~( t_4 - \epsilon~t_2) = - {\pi \over
K}~(K' + v),
}
\eqn\bctwo{\epsilon^2= { 4 K q \over i \pi\xi}~
{\theta_1\big({i \pi v \over K} ,q\big)
\over \theta'_1\big(0,q\big)},}
\eqn\bcthree{{2 t_4 \over \epsilon^2}-1 =
-{1\over 4}\zeta
          +\xi\bigl[{1\over 4~\sn(v,k')}\Upsilon+\Xi\bigr],
}
where we have defined
\eqn\defs{\eqalign{
\xi=&\sqrt{(a-c)(b-d)},\quad\zeta=a+b+c+d,\cr
\Upsilon=&
 3~\cn(v,k')~\dn(v,k')-
{\dn(v,k')\over\cn(v,k')}-{\cn(v,k')\over\dn(v,k')}
\quad{\rm and}\cr
\Xi=&{\pi\over 2K}+E(v,k')+\bigl({E\over K}-1\bigr)~v.}
}
$K$ and $K'$ are the complete elliptic integrals of the first kind
with respective moduli $k$ and $k'=\sqrt{1-k^2}$. $E$ is the complete
elliptic integral of the second kind with modulus $k$, $E(v,k')$ is
the
incomplete elliptic integral of the second kind with argument $v$ and
modulus $k'$ and $\sn$, $\cn$ and $\dn$ are the Jacobi Elliptic
functions. The nome $q$ is defined by
\eqn\defq{
q=e^{- \pi {K' \over K}}.
}
Finally $k$ and $v$ are defined in terms of the cut points by
\eqn\defkv{
k^2={(a-b)(c-d) \over (a-c)(b-d)}, \quad
      v=\sn^{-1}\bigl(\sqrt{{a-c \over a-d}},k'\bigr).
}
The final boundary condition will come from the normalisation of the
density.

The density $\rho(h)$ can be obtained from $2F(h)+H(h)$ by using the
saddle point equation $2F(h)+\cut H(h)=-\ln h$ and the fact that the
resolvent \res\ for the Young tableau can be written as $H(h)=\cut
H(h) \mp
i\pi\rho(h)$.  We thus perform the integrals in \cbbainfd\ and, after
using the first boundary condition and an identity between elliptic
functions\foot{ For this and many other relations between Jacobi's
elliptic functions and theta functions useful for performing the
calculations of this section see e.g.~\BYRD,\LAWD.}, we obtain

\eqn\density{\rho(h)=
{u \over K} - {i \over \pi}~\ln \Bigg[
{\theta_4\big({\pi \over 2 K} ( u - i v),q \big)
\over
 \theta_4\big({\pi \over 2 K} ( u + i v),q \big)} \Bigg], }
where $v$ is defined in \defkv\ and $u$ is defined by
\eqn\defu{
u=\sn^{-1}\bigl(\sqrt{{(a-h)(b-d) \over (a-b)(h-d)}},k\bigr).}
Integrating $\rho(h)$ from $b$ to $a$ and equating the answer to
$1-b$
to ensure that the density is normalized to $1$ (the flat portion
from $0$
to $b$ gives a contribution $b$), leads to the final boundary
condition:
\eqn\bcfour{
t_4={ -2 K i q \over \pi^2}~
{\theta_1\big({i \pi v \over K} ,q\big)
\over \theta'_1\big(0,q\big)}\bigl[-E+K\bigl(k'^2\sn^2(v,k')
  +{\Upsilon\,\Xi\over \sn(v,k')}+2\,\Xi^2\bigr)\bigr].
}
The elliptic modulus $k$ and the parameter $v$ are fixed in terms of
the coupling constants $t_2$, $t_4$ and $\epsilon$, by equations
\bcone\ and \bcfour\ . It is natural to use the parameter $\beta$
defined by $\beta^2={\epsilon t_2\over t_4}$ so that the elliptic
parameters
$k$ and $v$ are determined through the two parameters $t_4$ and
$\beta$.
Through \defkv\ $k$ and $v$ fix the two
independent ratios of differences between cut points. The scale and
position of the cut points are then fixed by equations
\bctwo\ and \bcthree\ .

We now concentrate on the free energy. We choose the
definitions for $t_2$, $t_4$ and $\epsilon$ introduced in section 2:
\eqn\ttetlb{
t_2=\sqrt{\lambda}t,\quad t_4=\lambda,\quad
                   \epsilon=\sqrt{\lambda}{\beta^2\over t},
}
which implies that
\eqn\ttq{
t_{2q}=\lambda^{{q\over 2}}\Bigl({\beta^2\over t}\Bigr)^{(q-2)}
\quad {\rm for } \quad q\geq 2.
}
This choice is motivated by several reasons. Firstly we want to have
a
way of measuring the area of the surfaces. The factors of $\lambda$
in
the coupling constants have thus been chosen in order that the power
of $\lambda$ counts the number of squares $A$
making up the surface (see eq.\PART). Each
corner of a vertex corresponds to one of the four corners of a square
and and so is weighted by $\lambda^{1\over 4}$. A vertex of $2q$ legs
(where $2q$ squares meet) is thus weighted with $\lambda^{{q\over
2}}$. The definition of $\epsilon$ is such that ${t_2\epsilon\over
t_4}=\beta^2$ as before, so that the elliptic parameters $k$ and $v$
are defined in terms of $\lambda$ and $\beta$. Finally, for surfaces
of spherical topology, each negative curvature defect is compensated
for by a precise number of positive curvature defects. In particular
a
negative curvature defect $t_{2q}$ is compensated by $q-2$ positive
curvature defects $t_2$ giving an overall factor of
$\beta^{2(q-2)}$. The power of $t$ will thus correspond to the
number of excess positive curvature defects. For a surface of
spherical topology which takes precisely four extra $t_2$ defects to
close the surface we would therefore expect a factor $t^4$ (see
\PART).

We now use a standard formula from matrix models. Denoting the free
energy by ${\cal Z}(t,\lambda,\beta)$, $Z=e^{-N^2{\cal Z}}$,
we have
\eqn\dlfem{
{\partial\over\partial\lambda}{\cal Z}(t,\lambda,\beta)=
      {1\over 4\lambda}(\langle{1\over N}\Tr M^2\rangle -1).
}
This identity corresponds to
grabbing hold of one of the propagators of a free energy
diagram. From equation \trMtwoq\ we see that
\eqn\mmh{
\langle{1\over N}\Tr M^2\rangle={1\over 2}+\langle h\rangle.
}
where $\langle h \rangle = \int dh~\rho(h)~h$.
Using the density \density\ to calculate $\langle h\rangle$ we
obtain,
after a long calculation,
\eqn\dlfe{\eqalign{
{\partial\over\partial\lambda}{\cal Z}=
{\xi^2\over 4\lambda}\Biggl[{Kk'^2\sn^2\over
8\pi~\cn~\dn}&(1-k'^2\sn^4)
-{k'^2\sn^2\over 4}-{1\over 2\pi}(K\,\Xi-{\pi\over
2})(\dn^2+k'^2\cn^2)\cr
&+{3\,\Upsilon\,\Phi\over 8\pi~\sn^2}
-{\Phi^2\over 2\pi^2\sn^2}-{K\,\Upsilon\over 4\pi~\sn}(1-\dn^2)
+{\Phi\,\Xi\over 2\pi~\sn}+{\Omega\,\Phi\over\pi~\sn}-{\Omega^2\over
2}
\Biggr],}
}
where
\eqn\defop{
\Omega={K\over\pi}(1-\dn^2+2\,\Xi^2)-\,\Xi\quad {\rm and}\quad
\Phi=\sn\, E-\Upsilon\bigl(K\,\Xi-{\pi\over 2}\bigr),
}
and for compactness we have denoted $\sn(v,k')$ by $\sn$ and
similarly for
$\cn$ and $\dn$.
Everything inside the square brackets of this expression is a
function
of the two elliptic parameters $k$ and $v$, in other words a
function of $\lambda$ and $\beta$. The $\xi$ outside the square
bracket
can be expressed from \bctwo\ and \ttetlb\ by
\eqn\xitt{
\xi={4 t^2 K q \over i \pi\beta^4\lambda}~
{\theta_1\big({i \pi v \over K} ,q\big)
\over \theta'_1\big(0,q\big)}.
}
We thus see that the free energy is written as a function of
$\lambda$
and $\beta$ times a factor $t^4$ (see \PART).
As discussed above, the
factor of $t^4$ corresponds to the four extra $t_2$ defects
needed to close the surface into the topology of a sphere.

Since the solution is somewhat
complicated, it is essential to check it carefully.
As a first check we investigate two limiting cases.
By setting $t_2=\epsilon=\sqrt{\lambda}$ and $t_4=\lambda$
(or equivalently $t=\beta=1$) we
correctly recover the solution found in \KSWI\ for the model with the
external matrices $A=B=\lambda^{{1\over 4}}{\cal J}$ where ${1\over
N}\Tr[{\cal J}^q]$ equals one for $q$ even and zero otherwise (see
equations (5.8)$\rightarrow$(5.11) of \KSWI\ ). Taking the other
limit
and setting $\epsilon=0$ we correctly recover the solution for the
flat space limit found in \KSWII . Both these limits are somewhat
subtle to extract since they correspond to singular points in the
boundary conditions, and thus have to be derived by a careful
taking of limits. We elaborate on this in appendix 2.~where further
nontrivial checks are also discussed.

\newsec{Extraction of physical results}

We are now in a position to derive the physical results announced in
section 2.  The boundary conditions contain sufficient information to
find the critical values of the couplings $\lambda$ and $\beta$
corresponding to the continuum limit (the elliptic parameters $k$ and
$v$ do not depend on the coupling $t$ which is therefore not fixed by
the continuum limit). The most direct way of finding the continuum
limit, however, is from the expression for the density \density . The
first term of the density, ${u\over K}$, is always positive on the
interval $(b,a)$. However, for a certain range of values of $v$, the
second term in \density\ can be negative. The continuum limit
corresponds to the point in $(q,v)$ parameter space where the density
is no longer positive. Specifically\foot{This is very
similar to ordinary matrix models.
There the point in coupling constant space at which the density of
{\it eigenvalues} becomes flat at its end point corresponds to the
continuum limit.}
we need to find the point in
$(q,v)$ space where the density becomes exactly flat at the end point
$a$. Taking the derivative w.r.t.~$h$ of the density at
the end point $a$ and setting it equal to $0$ gives
\eqn\qvcont{
1={\cn(v,k')\over \sn(v,k')~\dn(v,k')}\,\,\Xi,
}
where $\Xi$ is defined in \defs\ .  This equation defines a line in
$(q,v)$ space. Substituting \qvcont\ into \bcfour\ , a point on this
line can be related to $\lambda$ by
\eqn\tfocont{
\lambda=-{2Kiq\over\pi^2}{\theta_1({i\pi v\over
K},q)\over\theta_1'(0,q)}
                \Bigl[K{\dn^2(v,k')\over\cn^2(v,k')}-E\Bigr],
}
and $\beta$ by
\eqn\epttcont{
\beta^2= 1-{1\over \lambda}e^{-{\pi\over K}(K'+v)}.
}
Equations \qvcont , \tfocont\ and \epttcont\ define the continuum
line
in the parameter space $(\beta, \lambda)$.
A plot of this line is shown in Fig.~4.3.

\vskip 20pt
\hskip 50pt \epsfxsize=250pt\epsfbox{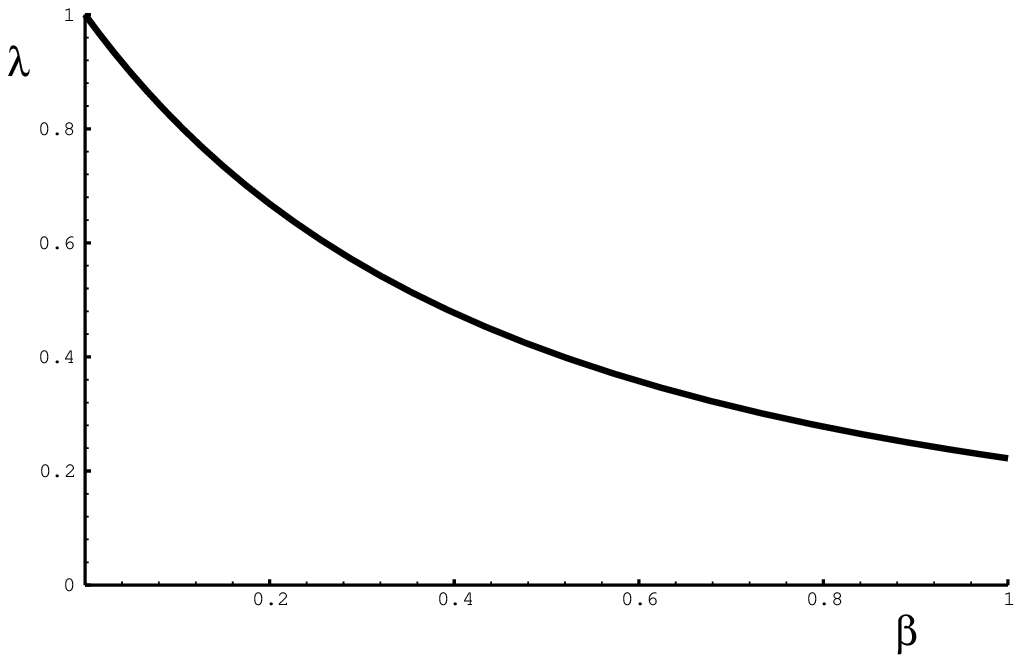}

\centerline{{\rm Fig.~4.3} The critical line in the $\beta$,
$\lambda$
plane}
\vskip 30pt

\hskip -19pt The point $\beta=1$
($\lambda={2\over 9}$) corresponds to the model
$A=B=\lambda^{{1\over 4}}{\cal J}$ mentioned above and solved in
\KSWI .
The point $\beta=0$ ($\lambda=1$)
corresponds to the flat space limit studied in \KSWII .  The
continuum
line corresponds to a singularity in the mapping between the coupling
constants $\beta$ and $\lambda$ and the parameters $q$ and $v$
through which physical quantities are expressed. The type of
singularity determines the physical regime of the continuum limit. It
is easiest to investigate the behaviour of the singularity about the
limiting points discussed above. The limit $\beta=1$ was discussed in
\KSWI\ , where a standard square root singularity was found,
corresponding to the pure gravity regime. We thus concentrate on the
limit $\beta\rightarrow 0$ and $\lambda\rightarrow 1$.

This limit corresponds to $q\rightarrow 1$. We can study it most
simply by expressing all the dependence on
the nome $q$ through the dual parameter $q'$, defined through
\eqn\defqp{
q'=e^{-\pi {K\over K'}}=e^{-{\pi^2\over \ln q}}.
}
We then discard all exponentially suppressed terms, i.e.~all powers
of
$q'$. This is not the same as setting $q'=0$ since we keep terms of
order $\ln q'$, in particular $K=-{\pi^2\over 2\ln q'}(1+{\cal
O}(q'))$. For $q$ close to one this is a good
approximation\foot{Numerically the computation of elliptic functions
is performed by writing them as a power series in either $q$ or $q'$,
whichever is smaller. The power series converge so rapidly that,
even for a generic point, the first few terms are sufficient to give
a
high degree of accuracy.}.  In this limit the boundary conditions
fixing $v$ and the nome $q$ are given by
\eqn\bcsqone{\eqalign{
{1\over 1-\beta^2}=&e^{{2zv\over\pi}}
                     \Bigl[\bigl(1+{2v\over\pi}\bigr)\cos 2v+
                    {1\over\pi}\bigl(z+{2vz\over\pi}-1\bigr)\sin
2v\Bigr],\cr
q=&\bigl[\lambda (1-\beta^2)\bigr]^{\bigl(1+{2v\over\pi}\bigr)},}
}
where we have defined $z=-\ln(\lambda(1-\beta^2))$.
Setting the derivative (w.r.t.~v) of the first of the above equations
to zero gives the critical line about the limiting point where the
surface flattens:
\eqn\critline{
\tan v_c={z\over\pi}.
}
This could also have been obtained by discarding all powers of $q'$
in
\qvcont\ . To investigate the type of singularity on the critical
line
we expand the r.h.s.~of the first equation in \bcsqone\ about the
critical value of $v$:
\eqn\critexp{
{1\over 1-\beta^2}=a_0+a_2(v-v_c)^2+{\cal O}\bigl((v-v_c)^3\bigr).
}
The coefficient $a_0$ is given by evaluating \bcsqone\ at $v_c$.
The coefficient $a_2$
is the second derivative (w.r.t $v$) of the first equation of
\bcsqone , also evaluated at $v_c$. It is easy to demonstrate that
$a_2$ cannot be zero for positive, real coupling coefficients. We
thus
find that the singularity on the critical line is a square root
corresponding to pure, two dimensional gravity. A more rigorous study
would analyse the singularity of boundary conditions \bcone\ and
\bcfour\ at a generic point along the critical line. We have examined
this numerically and find a square root singularity for the full
length of the critical line. {\it We can thus conclude that for all
finite
$\beta$ we are in the regime of pure, two dimensional gravity. }

We now proceed to extract the physical mechanism of flattening
which requires the limit $\beta\rightarrow 0$ (corresponding to
$v\rightarrow 0$). Expanding out
\bcsqone\ to lowest order in $v$, $z$ and $\beta^2$ we find
\eqn\vqlim{
v={\beta\over\sqrt{2}}\bigl(x-\sqrt{x^2-1}\bigr)\quad
{\rm and }\quad q=\lambda .
}
where we have introduced a scaling parameter $x$ defined by
\eqn\defx{
x=1+{\sqrt{2}\over\pi}{\mu\over\beta}\quad{\rm with}\quad
\mu=\lambda_c-\lambda\quad{\rm and}\quad
\lambda_c=1-{\pi\beta\over\sqrt{2}}+{\cal O}(\beta^2).
}
Note that this is a very natural, dimensionless scaling parameter;
$\mu$ is the continuum cosmological constant with dimension of
inverse
area and $\beta$ controls the number of curvature defects per unit
area (and thus also has dimension of inverse area).

Expressing the elliptic functions in \dlfe\ in terms of the dual nome
$q'$, dropping all powers of $q'$ and keeping only the leading order
contribution in powers of $v$, we arrive at the free energy about the
flattening transition point:
\eqn\feflat{
{\cal Z}(t,\lambda,\beta)=
{4 t^4\over 15\beta^2}\bigl[x^6-{5\over 2}x^4+{15\over 8}x^2-{5\over
16}
-x(x^2-1)^{5/2}\bigr].
}
This is the continuum scaling function corresponding to the
flattening
transition. For $x$ close to $1$, in other words for cosmological
constant $\mu$ much less than $\beta$, there is the standard
$\mu^{5\over 2}$ singularity characteristic of pure gravity. For
$x$ very large, corresponding to very small $\beta$ the free energy
has the singularity $\mu^{-2}$. {}From \feflat\ we can also
Laplace-transform
to the free
energy for fixed area
\eqn\fefiA{\eqalign{
{\cal Z}(t,A,\beta)=&
{8 t^4\over\pi^2 \beta^4 A^3}
  \Biggl[-\Bigl({48\over y^3}+{8\over y}\Bigr)I_1(y)
   +\Bigl({24\over y^2}+1 \Bigr)I_0(y)\Biggr]
               \quad{\rm with}\quad y={\pi\over\sqrt{2}}\beta A,\cr
\sim&\cases 
{t^4 \beta^{-{9 \over 2}} 
e^{{\pi\over\sqrt{2}}\beta A}A^{-{7\over 2}},&
                                 for $A\rightarrow\infty$\cr
           t^4 A,&for $A\rightarrow 0$}}
}
where $I_0$ and $I_1$ are modified Bessel functions. The limiting
behaviours for large and small areas are also shown.  For surfaces
whose area, $A$, is large compared to $\beta$ the exponent $-{7\over
2}$ signals that we are in the regime of pure 2D gravity. For
surfaces
of very small area $A$ there is a linear dependence on the area. The
function ${\cal Z}(A,\beta,t)$ smoothly interpolates between these
two
limiting behaviours. {\it There is no phase transition separating
flat
space from pure gravity}.

\newsec{Conclusions and open issues}

In conclusion, our result strongly suggests that
$R^2$ gravity \cont\ does not seem to exist as
a {\it continuum} theory. The discretisation of the term
${1 \over \beta_0}~R^2$ by the same principle that
consistently defines the path integral \cont\ for
$\beta_0 = \infty$ (i.e.~ordinary 2D gravity) does not
lead to any new fixpoints, indicating the {\it non-perturbative}
irrelevance of the dimensionful higher derivative terms of \cont.
Of course one might object that our particular (but generic!)
discretisation is too naive. Indeed, it would be gratifying
to prove the universality of our result by choosing different
weights \weights\ for the curvature terms in our model.
In principle, as we hope we have convinced the reader, this
can be checked with the help of our formalism.
It is not entirely excluded that fine-tuning the weights 
\weights\ in a subtle
way might result in a smooth phase of gravity (for some
(non-rigorous, as we believe) arguments in favor of this
hypothesis see \BUR). However,
it is equally evident that any further calculations
will be very tedious unless it proves possible to simplify
our method in a significant fashion.
We should also point out that the fact that the curvature
in the present type of discrete models is always necessarily
``quantised'' in units of $\pi$ could be an a priori stumbling block
for reaching a smooth phase of gravity. Could one find
discrete models of gravity that allow for soft, slowly varying
curvature fluctuations over many lattice spacings?

The present work presents the first rigorous step towards a
full understanding of these issues. It would be very interesting
to gain deeper insights by simplifying and extending
our novel discrete approach.

One might also attempt to describe a model like ours
directly in the continuum.
One could start with
the conformal metric of a flat surface with localised curvature
defects. It can be represented locally as
$g_{ab} = \delta_{ab}~e^{\varphi(z)}$ with
\eqn\metr{  \varphi(z)=  \sum_{j=1}^M~R_j~\ln(z-z_j)^2,   }
where $R_j=-1,1$. Symbolically,
the partition function might be written as
\eqn\partf{ Z(\mu,\b)=
\sum_M~\beta^M~\int d [z_1,...,z_M]
{}~e^{-\mu \int d^2 z~\sqrt{\det g(z)} } .}
Here we introduced the fugacity of curvature defects $\b$ instead of
the explicit $R^2$-term in the action. It serves the same purpose:
for
$\beta \rightarrow 0$
we arrive at the completely flat metric, whereas for $\b \sim 1$ the
system should show the behaviour of pure quantum gravity, at least in
the infrared domain. We retained the notation $\b$ to denote the
parameter playing a role similar to the
$R^2$ coupling in the above discrete model.

This formulation resembles a little bit the two-dimensional Coulomb
gas
problem. However, the measure of integration $d [z_1,...,z_M]$ of
the positions of the curvature defects is a complicated object: it
should take into account the topology of the surface and the
existence
of zero modes (the action does not depend on some directions in the
space of the $z_i$). It would be very interesting to make
this direct continuum formulation more precise.

Another interesting issue is the role of exponential corrections
appearing due to the structure of elliptic functions.
In fact all physical quantities, such as 
${\cal Z}(t,\lambda,\beta)$ (see \PART)),
contain exponentially small terms in the limit 
$\lambda \rightarrow \lambda_c$ and $\beta \rightarrow 0$, 
thus leading to an essential
singularity at $\beta=0$, $\lambda=1$. 
These terms are precisely the powers of $q'$ that we discarded to
arrive at the scaling function \feflat\ of section 5.
One can obtain the first
correction of this type in e.g.~the free energy 
${\bar f}(\beta)$ per unit area
in the thermodynamical limit $\lambda=\lambda_c$:
\eqn\expc{
{\cal Z}(t,A,\beta)~\sim~{t^4\over\beta^{9\over 2}}~
e^{{\bar f}(\beta) A}~A^{-{7 \over 2}}
\quad {\rm with} \quad
{\bar f}(\beta)= 
{\pi \over \sqrt{2}} \beta \Bigl[(1+\dots ) 
+ e^{-{\pi \sqrt{2}\over\beta}}(4+\dots) \Bigr] }
where 
${\bar f}(\beta)=\lim_{A\rightarrow\infty}
{1\over A}\ln {\cal Z}(t,A,\beta) =\lambda_c$ 
and the dots denote terms of order $\beta^3$ and higher.

These exponential terms are likely to be lattice artifacts. 
They emerge even in the simplest calculation for the flat
closed quadrangulation with four positive curvature defects,
where they appear as discrete corrections to the approximation of 
elliptic sums by integrals close to the continuum limit.

On the other hand, formula \expc\ 
corresponds to the critical free energy
as a function of the curvature fugacity $\beta$ 
(i.e.~we have already taken
the continuum limit).
It is possible that the exponential terms might be 
corrections relevant for the
statistical mechanics of random lattices at long distances 
(of order 
$1 \over \beta$) rather than for continuous 2D-gravity.

However, the most interesting and rewarding extension of our methods
and ideas would be their generalisation to allow for the coupling
of matter to discrete two-dimensional $R^2$ gravity. 
It would lead to an entirely novel approach to two-dimensional
physics, unifying theories in the absence and presence of
two-dimensional quantum gravity.

\vskip 30pt
\hskip -19pt{\bf Acknowledgements}

We would like to thank E.~Br\'ezin, I.~Kostov, D.~Nelson, A.~Polyakov
and
A.~B. Zamolodchikov for useful discussions.

\appendix{1}{General case: Large $N$ limit of Schur characters}

The two equations, \grav\ , in which there are three non-zero
couplings,
$t_2$, $t_4$ and $t_6$, can be reduced to a single integral equation
for the density. The derivation we present below can be
generalised in an obvious manner to an arbitrary finite set of
non-zero couplings. Other models with different saddlepoint equations
can also be treated with only minor modifications.

The initial steps are very similar to the start of section 4, to
which
we refer the reader for more details. We start from the equation
\eqn\hGtfs{
h-1={t_2\over G}+{t_4\over G^2}+{t_6\over G^3}+
   {\rm\ positive\ powers\ of\ } G.}
and deduce (by identical reasoning to section 4) that the physical
sheet of the function $G(h)$ and all the sheets attached to it by
the cut of $e^{F(h)}$ have a cubic structure. We further deduce that
the function $\ln G(h)$ has a logarithmic singularity starting on the
physical sheet at the point $h=b$, and, by looking at the large $h$
limit of \hGtfs\ (inverting \hGtfs\ to lowest order in ${1\over h}$
gives
$\ln G(h)={1\over 3}\ln ({t_6\over h})+{\cal O}(h^{-{1\over 3}})$ )
deduce
that the logarithmic cut goes off to infinity. We now study the
combination $2\ln G(h)-\tilde H(h)$ where $\tilde H(h)$ is defined in
\tildefh\ . We find it has the analytic structure shown in Fig.~A1

\vskip 20pt
$$\epsfbox{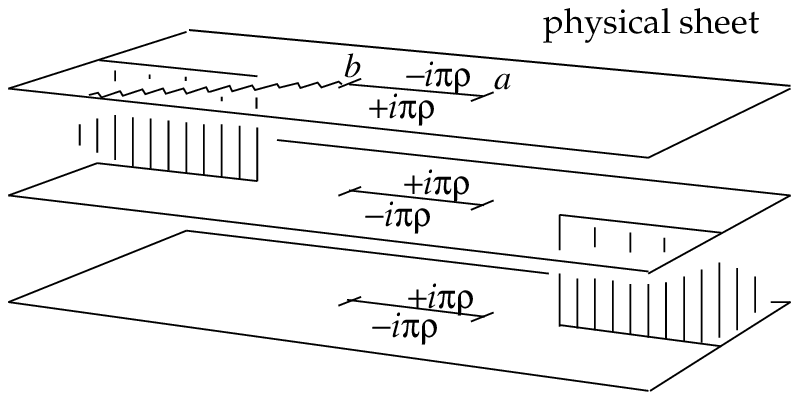}$$
\centerline{{\bf Fig.~A1} $2\ln G(h)-\tilde H(h)$ in the complex $h$
plane}
\vskip 30pt

\hskip -19pt
where the zig-zag line, starting at
$h=b$ on the physical sheet, represents the logarithmic cut. It also
has
a saddlepoint equation (trivially derived from \grav\ and \tildefh\ )
of
\eqn\tFmHsdpt{
\Re [2\ln G(h)-\tilde H(h)] = -\ln(h-b).
}
We now unwind the three sheets of Fig.~A1 by parametrising $h$ as a
cubic polynomial of a new parameter $z$
\eqn\hz{
h=z^3-\alpha z +\gamma.
}
The coefficients $\alpha$ and $\gamma$ are such that the cut points
of
$h$ as a function of $z$ are identical to the cubic cut points of
Fig.~A1. The three sheets of Fig.~A1 (defined in the complex $h$
plane) then unwind to become a single sheet with three cuts (plus a
logarithmic cut) in the complex $z$ plane (see Fig.~A2).

\vskip 20pt
$$\epsfbox{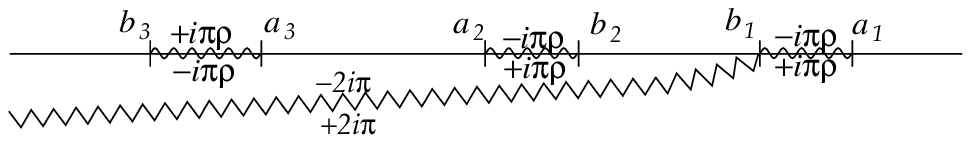}$$
\centerline{{\bf Fig.~A2} $2\ln G(h(z))-\tilde H(h(z))$ in the
                                                complex $z$ plane}
\vskip 30pt

\hskip -19pt Defining
\eqn\fz{
{\cal F}(z)=2\ln G(h(z))-\tilde H(h(z))
-2\ln({t_6^{1\over 3}\over z-b_1}),
}
we see that ${\cal F}(z)$ is an analytic function in the complex $z$
plane with three cuts corresponding to the images of the cut of the
density. Furthermore, we see from the large $h$ limits of $\ln G(h)$
and $\tilde H(h)$, given by
$\ln G(h)={1\over 3}\ln ({t_6\over h})+{\cal O}(h^{-{1\over 3}})$ and
$\tilde H(h)={\cal O}({1\over h})$, that the function ${\cal F}(z)$
behaves as ${\cal O}({1\over z})$ for large $z$. The function
${\cal F}(z)$ can thus be generated entirely from the discontinuities
of its cuts:
\eqn\fzrho{
{\cal F}(z)=\int_{b_1}^{a_1}\,ds{\rho(h(s))\over z-s}
-\int_{b_2}^{a_2}\,ds{\rho(h(s))\over z-s}
-\int_{b_3}^{a_3}\,ds{\rho(h(s))\over z-s}.
}
We now use the identity
$(z_1-z_1')(z_1-z_2')(z_1-z_3')=h-h'=(z_1-z_1')(z_2-z_1')(z_3-z_1')$,
where the $z_i$ are the three roots of \hz\ and the $z_i'$ are the
three roots of \hz\ with $h$ set to $h'$. Along with
equation \fz\ , this leads to
a saddle point equation in the complex $z$ plane:
\eqn\fsdpt{
\barint_{b_1}^{a_1}\,ds\,\rho(h(s))\Bigl[{1\over z-s}
-{1\over z_2(z)-s}
-{1\over z_3(z)-s}\Bigr]
=\ln\bigl[{(z-b_1)\over (z-b_2)(z-b_3)}\Bigr]-{2\over 3}\ln t_6,
}
where $z$, $z_2$ and $z_3$ are the three roots of \hz\ . $z_2$
and $z_3$ are thus functions of the first root $z$. With the
parametrisation of \hz\ they would be given by
$z_{2(3)}(z)={z\over 2}\pm\sqrt{\alpha-{3z^2\over 4}}$.

The generalisation to more than three weights is obvious. However,
the
functions $\rho$ that satisfy such an equation, even for the case of
just three weights, involve functions more general than elliptic and
probably more general even than hyperelliptic.

As a final comment, notice that by unwinding the analytic structure of
the function $\ln G(h)$ using the mapping \hz\ as in the discussion
above we arrive at a closed expression for the function 
$\,\ln G(h)=F(h)+H(h)$:
\eqn\Grho{
\ln G(h)= \int_{b_1}^{a_1}\,{ds\,\rho(h(s))\over z-s}
+\ln\bigl({t_6^{1\over 3}\over z-b_1}\bigr),
}
where $z$ is related to $h$ by \hz . Given a representation, specified
in the large $N$ limit by the density $\rho(h)$ (which then defines
$H(h)$ through \res ), we thus have a closed expression for the
logarithmic derivative of the character, $F(h)$, \defF .
The coupling $t_6$, is incorporated directly into the expression
\Grho , the two remaining couplings, $t_2$ and $t_4$, determine, 
through \hGexp , the coefficients $\alpha$ and $\gamma$ of the mapping
\hz . It is again trivial to generalize this result to more than three
weights.

\appendix{2}{Verification of the solution}

\subsec{Limit $\beta\rightarrow 1$}
This corresponds to
the model with $A=B=\lambda^{{1 \over 4}}{\cal J}$ 
studied in \KSWI\ . Setting
$t_2=\epsilon=\sqrt{\lambda}$ and $t_4=\lambda$ (or equivalently
$t=\beta=1$) the first boundary condition, \bcone , implies
immediately
that $K'\rightarrow\infty\Rightarrow k=0\Rightarrow c=d$. The values
of $E$, $K$, $E(v,k')$, $\sn(v,k')$, $\Xi$ and $\Upsilon$ are then
given by
\eqn\beonevals{
E=K={\pi\over 2},\quad E(v,k')=\sn(v,k')=1,\quad\Xi=2\quad{\rm
and}\quad
\Upsilon=-\sqrt{{a-c\over a-b}}-\sqrt{{a-b\over a-c},}.
}
where we have used the definition of $v$ in terms of the cut points,
\defkv , along with standard identities $\cn^2=1-\sn^2$ and
$\dn^2=1-k'^2\sn^2$ to find $\Upsilon$. After using \bctwo\ to
eliminate the theta functions in \bcfour , and also using \bcthree ,
we find that $c=d=0$. In other words the cut of F(h) disappears. We
also find the first condition fixing $a$ and $b$:
\eqn\beoneone{
\sigma-3\varphi-1=0\quad
{\rm where}\quad\sigma=\Bigl({\sqrt{a}+\sqrt{b}\over 2}\Bigr)^2\quad
{\rm and}\quad\varphi=\Bigl({\sqrt{a}-\sqrt{b}\over 2}\Bigr)^2.
}
The limit of the third boundary condition, \bctwo , is more
subtle. We shift the argument of the theta function using the
identity: $\theta_1(z,q)=qe^{-2iz}\theta_1({i\pi K'\over K}-z,q)$ and
carefully take the limit
${\pi\over 2K}(K'-v)\rightarrow\sinh^{-1}\sqrt{{b\over a-b}}$ to
find the final boundary condition
\eqn\beonetwo{
3\lambda^2\sigma^3-\sigma+1=0.
}
These were the conditions found in \KSWI\ leading to
$\lambda_c={2\over 9}$. The density \density\ can likewise be reduced
down to that found in \KSWI .

\subsec{Limit $\beta\rightarrow 0$ i.e. $\epsilon\rightarrow 0$}
This
limit recovers the flatspace solution found in \KSWII .  In this
limit $d\rightarrow -\infty$ and the density \density\ trivially
reduces to that found in \KSWII . To see that the boundary conditions
give the correct limit and also to check perturbatively that 
the next few
powers in $\beta$ correctly correspond to negative curvature
insertions we expanded the results in
powers of $v$. Expansions of elliptic functions can
be found in most mathematical tables. A comprehensive list is found
in
\BYRD . Below we give the expansion formula for the theta
function which is less common
\eqn\beOthexp{
{\theta_1\big({i \pi v \over K} ,q\big)\over \theta'_1\big(0,q\big)}
={i\pi\over K}v
   -{2i\pi\over 3K}\bigl(1+k'2-3{E\over K}\bigr)v^3+{\cal O}(v^5).
}
Expanding out \bcone\ and \bcfour\ leads to
\eqn\beOlamb{
v={K\over 2\pi}\beta^2 +{\cal O}(\beta^4)\quad\quad{\rm and}\quad
q=t_4-{t_4\beta^2 \over 2}+{\cal O}(\beta^4).
}
For
$\beta=0$ we recover the first boundary condition of \KSWII\ .  Using
the identities $(a-c)={\sn~\dn\over\cn}\xi$ and $\epsilon={\beta^2
t_4\over t_2}$ along with \beOlamb\ and \bctwo\ and working to lowest
order we find the second boundary condition of \KSWII :
\eqn\beOamc{
{t_2\over \sqrt{t_4}}={\pi\over K}\sqrt{a-c}.
}
A careful
expansion of \bcthree\ then leads to the final boundary condition
found in \KSWII :
\eqn\beOa{
a=1+{t_2^2\over \pi^2t_4}(K^2-EK).
}

A final check of the solution can be performed by expanding out the
expression for the free energy \dlfe\ in powers of $v$ and hence in
powers of $\beta$. To lowest order we correctly recovered the free
energy found in \KSWII . Expanding out to the first non-zero order in
$\beta$ we retrieved the order $\beta^2$ contribution. The result
matched exactly with the calculation of a single negative curvature
insertion of $t_6$, calculated in \KSWII .

\listrefs
\end